\documentclass{IEEEtran}
\usepackage{cite}
\usepackage{amsmath,amssymb,amsfonts}
\usepackage{algorithmic}
\usepackage{graphicx}
\usepackage{gensymb}
\usepackage{textcomp}
\usepackage{makecell}
\usepackage[caption=false,font=footnotesize]{subfig}
\def\BibTeX{{\rm B\kern-.05em{\sc i\kern-.025em b}\kern-.08em
    T\kern-.1667em\lower.7ex\hbox{E}\kern-.125emX}}
\begin{document}
\title{Low-Cost and Low-Weight Horn and Reflector Antennas through 3D Printing}
\author{Alejandro Rivera-Lavado, Luis-Enrique Garc\'ia-Mu\~noz, Gabriel Santamar\'ia Botello, and Daniel Segovia-Vargas
\thanks{This work has been financially supported by ``DiDaCTIC: Desarrollo de un sistema de comunicaciones inal\'ambrico en rango THz integrado de alta tasa de datos'', TEC2013-47753-C3, TEC2016-76997-C3-2-R CAM S2013/ICE-3004 ``DIFRAGEOS'' projects, ``Proyecto realizado con la Ayuda Fundaci\'on BBVA a Investigadores y Creadores Culturales 2016'' and ``Estancias de movilidad de profesores PRX16/00021'', ``C\'atedras de Excelencia Banco de Santander 2017''}
\thanks{Alejandro Rivera-Lavado, Luis-Enrique Garc\'ia-Mu\~noz, Gabriel Santamar\'ia Botello, and Daniel Segovia-Vargas are with Universidad Carlos III de Madrid. Departamento de Teor\'ia de la Se\~nal y Comunicaciones, Legan\'es, Madrid, Spain (e-mail: dani@tsc.uc3m.es)}}

\maketitle

\begin{abstract}
The revolution that wireless communication is suffering during the last decades has increased the need of antennas where performances such as lightweight, rapid manufacturing and portability run in parallel with their electric performance. The maker revolution in which 3D printers can help manufacturing customized, high performance, lightweight and rapid manufacturing antennas is a fact to be explored. This paper discusses the design, simulation and 3D printing of an off-axis reflector antenna. The proposed design is composed of a parabolic reflector and an X band circular horn antenna as a feed. By adding a dielectric rod waveguide (DRW) to the feed, a rotationally symmetric radiation patter, similar to the one with a corrugated horn, is achieved.
\end{abstract}


\section{Introduction}

Since the appearance of commercial 3D printers, its performance has been increasing while its unitary and operational costs keep decreasing \cite{ref:3DPIntro}. This makes them common tools in different areas, such as research and education, despite of its main weakness: the printing time.

During the last years the antenna community opened to this new technology in order to look for low-cost and low-weight performance in traditional antenna technology. Till now most contributions in the antenna field have been related to low profile antennas \cite{ref:LowProfileAntennas1, ref:LowProfileAntennas2,ref:LowProfileAntennas3}. There are not many contributions related to 3D manufacturing of electrically larger antennas \cite{ref:ElectricallyLargeAntennas1,ref:ElectricallyLargeAntennas2}. Indeed, when looking for electrically large and non-planar antennas working at microwave frequencies, it is difficult to find contributions since they would require the assembly of different large radiating structures such as horn and reflector antennas. This communication, from the electric point of view presents a reflector antenna directly fed through a horn antenna made through 3D printing technology. 

From the 3D printing (3DP) point of view, the 3DP methodology has made use of printer common filaments, such as polylactic acid (PLA). PLA has been characterized in several contributions \cite{ref:PLA1,ref:PLA2} for different frequency ranges and is especially appealing due to its low cost, easy to print and biodegradability. Furthermore, its electrical properties, both permittivity and losses, are well known even at microwave frequencies \cite{ref:PLA1,ref:PLA2}. Despite providing these suitable characteristics PLA suffers from the variability of its dielectric properties for RF designs \cite{ref:MCAllister1983, ref:Petosa1998, ref:Yeh2008} depending on the providers. 

This paper initially presents a setup made on 3DP and composed of a feeding horn antenna assembled with a reflector working in the X band from 8 to 11 GHz. In order to enhance the overall antenna gain and the radiation pattern symmetry, the classical horn antenna made on 3DP and metalized with a nickel paint has been modified with the inclusion of dielectric rod waveguide (DRW) also made on 3DP. The new DRW inserted in the 3DP horn antenna improves the illumination efficiency of the reflector and achieves radiation symmetry in both E and H plane. This horn design and the reflector are described in Section II. Section III describes the 3D printing manufacturing process. Section IV shows the measurement results. 

\section{System Design}
The proposed design has been conceived with two main objectives. First, to show an easier to manufacture alternative to corrugations for achieving rotational symmetry in the horn radiation pattern. This is achieved by placing a DRW inside the horn. Second, to show that additive technologies allow manufacturing of customizable reflector systems. 

\begin{figure}[h]
	\centerline{\includegraphics[width=\columnwidth]{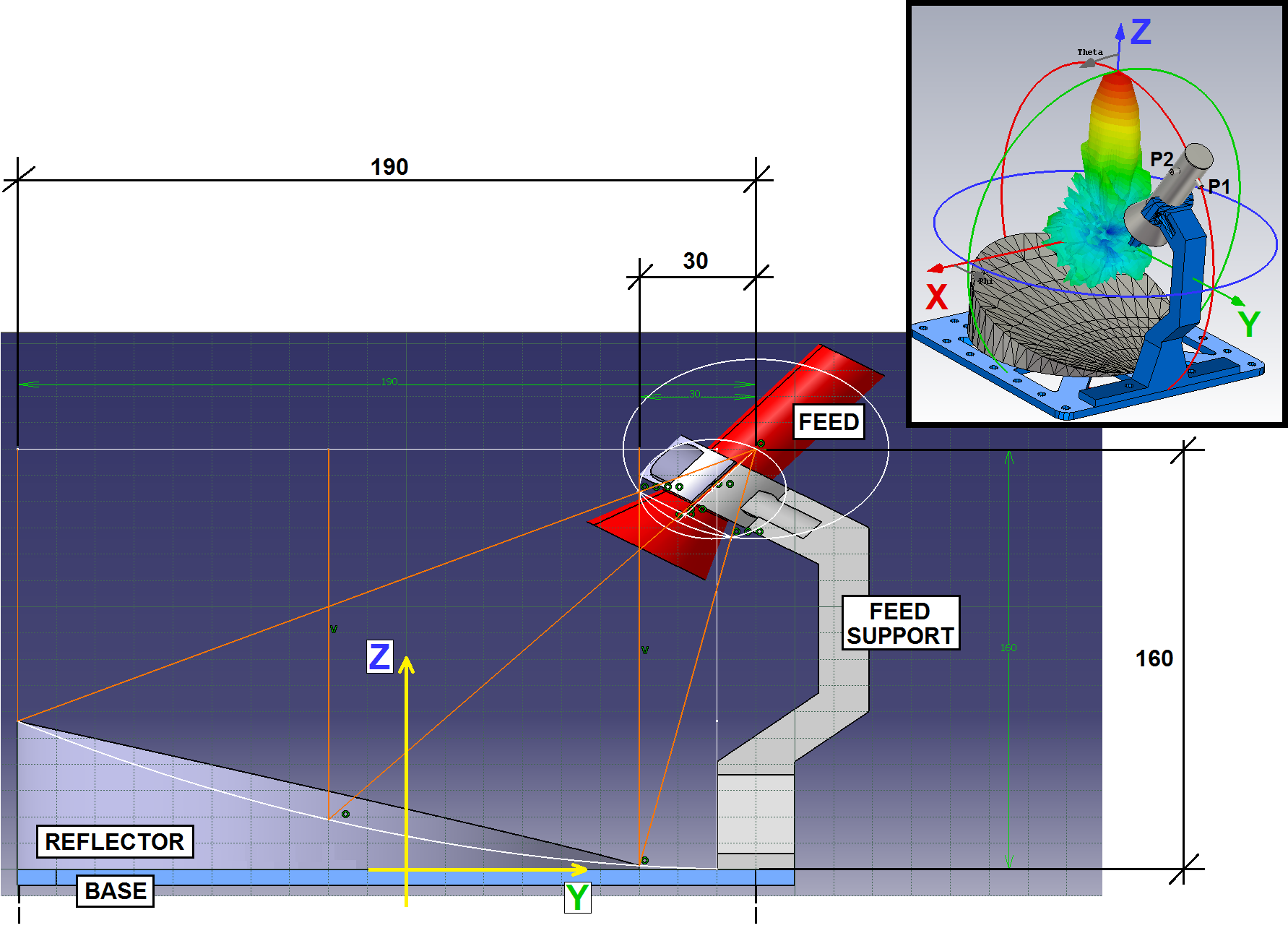}}
	\caption{Sketch of the system. The phase center of the feed (a circular horn is shown, red) is placed in the focal of the parabolic reflector (purple). The disc lays in a base (blue) on which the feed support (grey) is fixed. The coordinate axis center (yellow) corresponds to the base center. The inset shows an isometric view.}
	\label{fig:Sketch}
\end{figure}

Fig. \ref{fig:Sketch} shows a sketch of the design. The whole antenna is assembled on a base of 200x200 mm, whose center is placed on the origin. A feed support fixes a circular horn, so its phase center lies in the focal of the parabolic reflector. The horn has two SMA connectors, the first (P1) lying in the YZ plane and the second (P2) an orthogonal one. This allows its use as a dual linear polarization system; or a dual circular polarization one if an orthomode transducer is used. The use of Only results when exciting from the YZ plane (from now, E-plane) connector are given.

\subsection{Horn Feed}

\begin{figure}[h]
	\centerline{\includegraphics[width=\columnwidth]{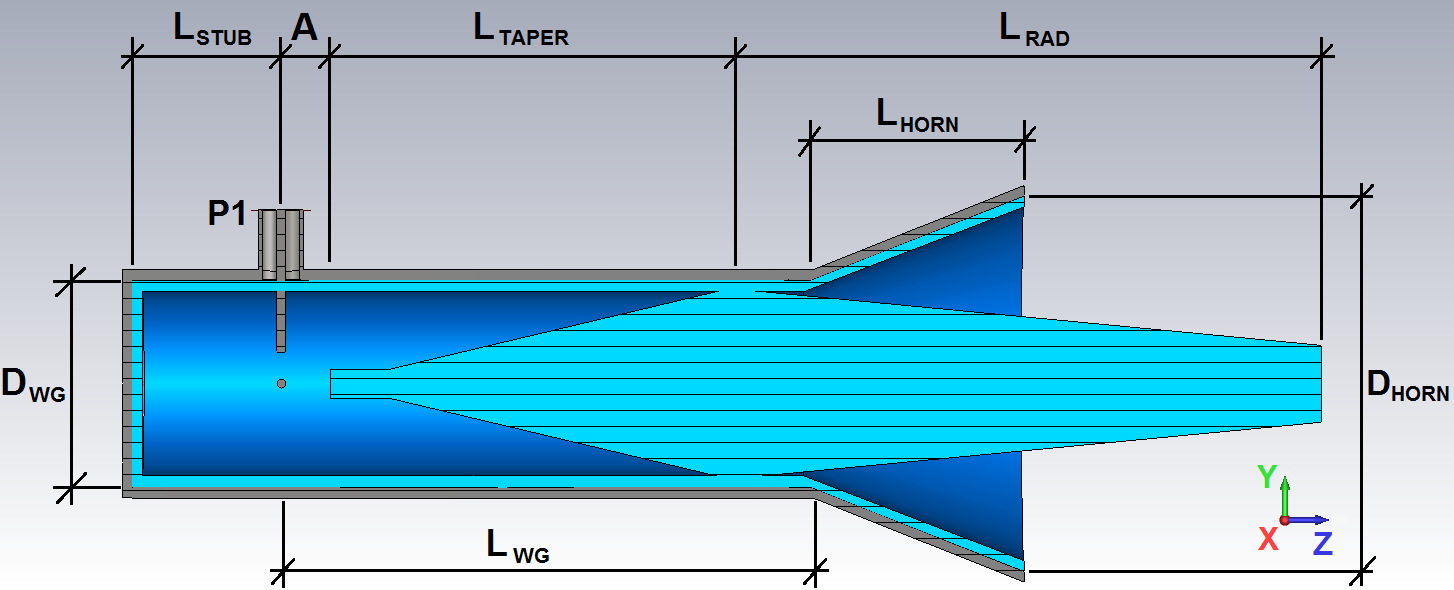}}
	\caption{Sketch of the feed. All dimensions are adjusted for working from 8 GHz to 11 GHz.}
	\label{fig:Sketch_HORN}
\end{figure}

Fig \ref{fig:Sketch_HORN} sketches the feed in the YZ plane. A circular horn with a DRW is made of PLA. Then, an external conductive paint layer is applied. The second coaxial port (P2) is not shown in this view. The DRW has two taper sections. The first one has a length of $L_\text{TAPER}=40$ mm for ensuring a proper matching of the field from the circular waveguide to the DRW. The second one, of length $L_\text{ROD}=60$ mm radiates the fields into the free space. Rod thickness is 3 mm. The DRW cross section dimensions are determined for ensuring a single mode regime at the working frequencies. The DRW modal chart can be obtained following the procedures described in \cite{ref:Yeh2008,ref:DRWModal1,ref:DRWModal2}. Two orthogonal DRW, one per polarization, are used.

The circular waveguide has a diameter $D_\text{WG}=21$ mm and a length $L_\text{WG}=54$ mm. The horn has a length $L_\text{HORN}=47$ mm and a diameter $D_\text{HORN}=38$ mm. $L_\text{STUB}$ is 15 mm and A is 5 mm. The PLA (blue) is 1 mm thick. 3D printing technologies allows the freedom of setting any of these parameters according to the requirements of the specific application. Conductive paint (grey) is assumed lossless for the following simulation.

\begin{figure}[h]
	\centerline{\includegraphics[width=\columnwidth]{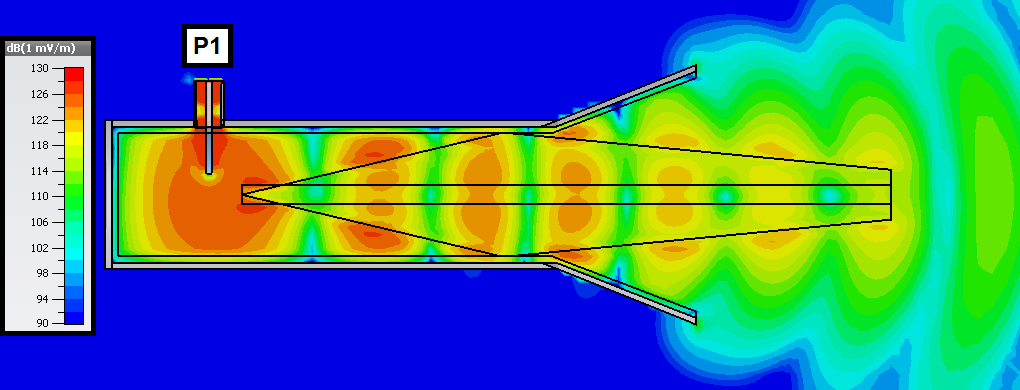}}
	\caption{Simulated E-field distribution on YZ plane at 9.5 GHz.}
	\label{fig:E-field_HORN}
\end{figure}

\begin{figure}[h]
	\centerline{\includegraphics[width=0.9\columnwidth]{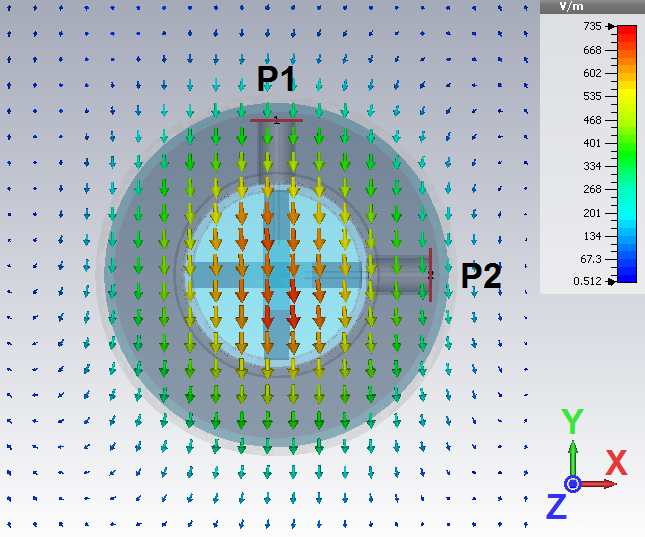}}
	\caption{Distribution of the radiated E-Field.}
	\label{fig:RadiatedMode}
\end{figure}

\begin{figure}[h]
	\centerline{\includegraphics[width=0.65\columnwidth]{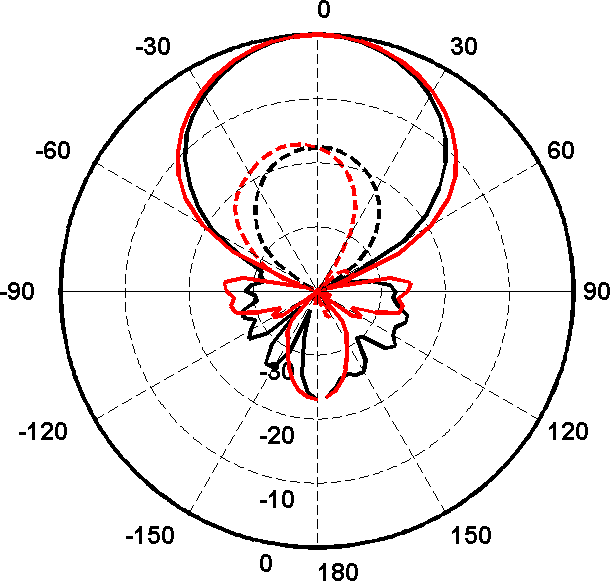}}
	\caption{Simulated radiation pattern for the circular horn. Both E-plane (black) and H-plane (red) co-polar (solid) and cross-polar (dashed) components are shown.}
	\label{fig:DR_HORN}
\end{figure}

The simulated E-field distribution in the E-plane is shown in Fig. \ref{fig:E-field_HORN}. The matching taper is slightly larger than the wavelength in the circular waveguide $\lambda_\text{WG}$, which avoids the presence of a standing wave\cite{ref:DRWGeneralov,ref:DRWPaper}. The radiation taper is around two wavelengths length. It is truncated since the bandwidth is limited by the high cutoff frequency of the circular waveguide\cite{ref:DRWPaper}. According to our simulations, no higher order modes are present, nor inside the horn nor in the DRW for frequencies below 11 GHz, which ensures a single beam radiation pattern\cite{ref:DRWPaper} from 8 to 11 GHz. By using a DRW, we achieve a field distribution similar to the $HE_{1,1}$ mode, as it can be seen in Fig. \ref{fig:RadiatedMode}. This increases the rotational symmetry of the radiation pattern (Fig. \ref{fig:DR_HORN})\cite{ref:rotationalSymmetryHe11}.

Fig. \ref{fig:DR_HORN} shows the radiation pattern at 9.5 GHz. According to our full-wave simulations, the use of the DRW taper increases the achieved gain from 9.1 dB to 12 dB. 

\subsection{Reflector}

\begin{figure}[h]
	\centerline{\includegraphics[width=0.92\columnwidth]{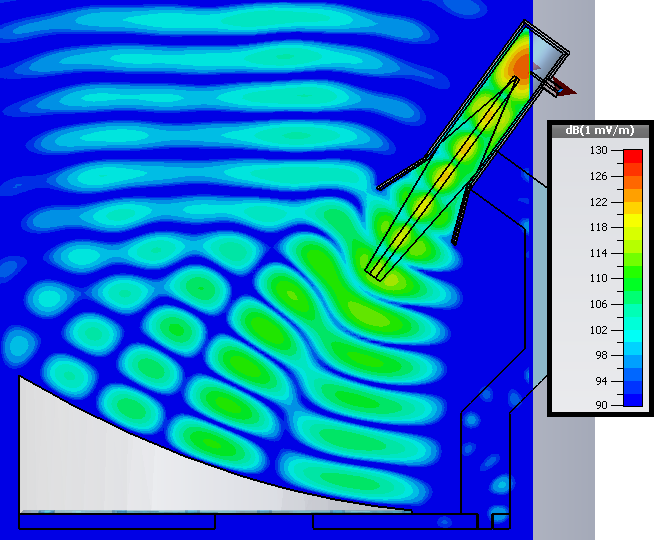}}
	\caption{Simulated E-field distribution on YZ plane at 9.5 GHz.}
	\label{fig:E-field_REFL}
\end{figure}

Considering the center of the antenna base as the origin of coordinates (see Fig. \ref{fig:Sketch}) the reflector surface can be defined as the part of the paraboloid

\begin{align*}
z&=\frac{(y+y_f)^2+x^2}{4 \cdot z_f}
\end{align*}

lying in the disc defined by the circle of radius $R=80$ mm and center $C_0(x,y,z)=(0,20,0)$ mm. The paraboloid focus is in $(x_f,y_f,z_f)=(0,90,160)$ mm and the vertex in $(x_v,y_v,z_v)=(0,90,0)$ mm. Simulated E-field distribution is shown in Fig. \ref{fig:E-field_REFL}. The simulated radiation pattern is shown in Section IV next to the measured one, for comparison purposes.

\section{Manufacturing}

All the parts were modeled using the full-wave solver \cite{ref:cst}. Then, the 3D models were exported in a .STL file. This format can be processed with the used slicing software \cite{ref:cura3D}. Finally, the generated GCODE is fed into the 3D printer in an SD card.

The design was manufactured using a dual extruder Sigma BCN3D 3D printer \cite{ref:BCN3D}. The printer can reach a layer resolution of $50 \mu$m, but $100$ $\mu$m was used instead as a good compromise between reflector surface quality and printing time. Nevertheless, the reflector surface was polished with a die grinder for ensuring a surface roughness below $\lambda/20=c/(20\cdot f_0)\approx150$ $\mu$m.

A fill density of 15\% and an external layer thickness of 0.6 mm was configured for the reflector. Both the base and the reflector were printed as a single piece, which took 200 grams of PLA and 15 hours.  The feed fill density is 100\%. The whole printing process for all the parts of the prototype took 25 hours and less than 300 grams of PLA (not including the sacrificial material). 3D printing is too time-consuming for mass production, but still more convenient than subtractive manufacturing for prototyping.

\begin{figure}[h]
	\centerline{\includegraphics[width=0.6\columnwidth]{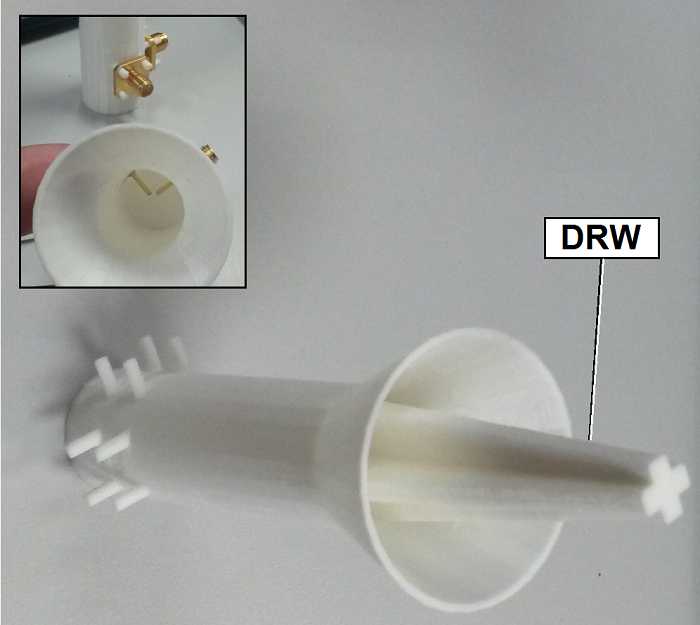}}
	\caption{Manufactured feed before painting with conductive ink. The horn and the DRW was 3D printed as a single piece. The inset shows a detail of the SMA connectors for dual polarization in a horn antenna without a DRW.}
	\label{fig:FeedHorn}
\end{figure}

The circular horn was printed with the DRW taper in a single piece (Fig. \ref{fig:FeedHorn}). The external surface was painted with a conductive spray paint. An extra layer of non-conductive paint was applied for protecting the conductive one. The same procedure was followed for the reflector surface.

A nickel-based conductive paint, which has higher losses than copper-based paints, was used for availability. In order to determine the losses, a second reflector was printed, polished and metalized using aluminum foil. Both reflectors were measured. The $S_{21}$ amplitude obtained with the aluminum foil metalized reflector when pointing the main beam to the reference antenna was 0.2 dB higher than the one obtained with the nickel paint metalization, with no further relevant differences among them.  

Eight PLA rods were included in the feed design for fitting the wall-mounted type SMA connectors, since they can be easily melted with a soldering iron (Fig. \ref{fig:FeedHorn}). The connectors are fixed in the last step of the horn manufacture.

\section{Measurement Results}

\begin{figure}[h]
	\centerline{\includegraphics[width=0.7\columnwidth]{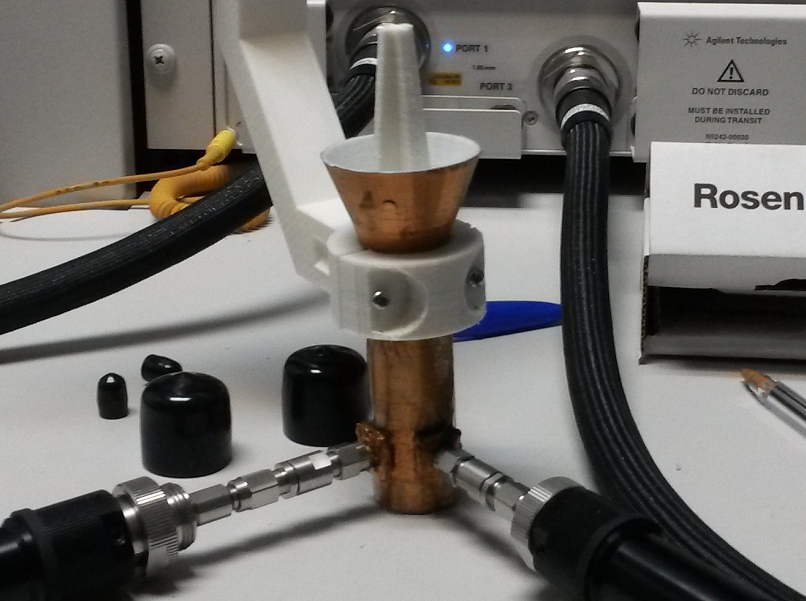}}
	\caption{S parameter characterization setup. The prototype was measured from 6 GHz to 12 GHz using a Keysight PNA-X.}
	\label{fig:Sxx_Setup}
\end{figure}

\begin{figure}[h]
	\centerline{\includegraphics[width=0.885\columnwidth]{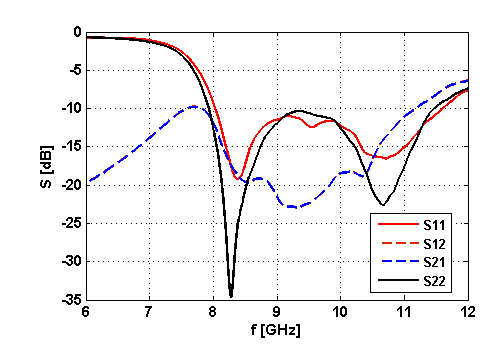}}
	\caption{Measured S parameters. A return loss greater than 10 dB is achieved from 8 to 11.5 GHz. Isolation is greater than 22 dB at 9.5 GHz.}
	\label{fig:Sxx}
\end{figure}

The prototype was fully characterized. First, the full 2x2 S parameter matrix was obtained from 6 to 12 GHz using a Keysight PNA-X (Fig. \ref{fig:Sxx_Setup}) and shown in Fig. \ref{fig:Sxx}. The feed horn achieves a return loss below 10 dB from 8 GHz to 11.5 GHz, which exceeds the intended range, from 8 to 11 GHz ($f_0=9.5$ GHz) for both polarizations.

\begin{figure}[h]
	\centerline{\includegraphics[width=0.7\columnwidth]{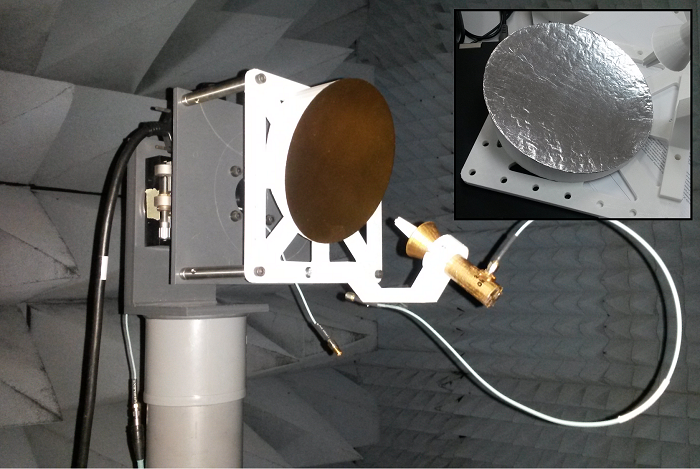}}
	\caption{The radiation pattern was measured in an anechoic chamber. The base of the reflector is designed for an easy integration to this measurement setup. The inset shows the reflector metalized with aluminum foil.}
	\label{fig:DR_Setup}
\end{figure}

\begin{figure}[h]
	\centering
	\mbox{\subfloat[]{\label{fig:DR_SIM} \includegraphics[width=0.45\columnwidth]{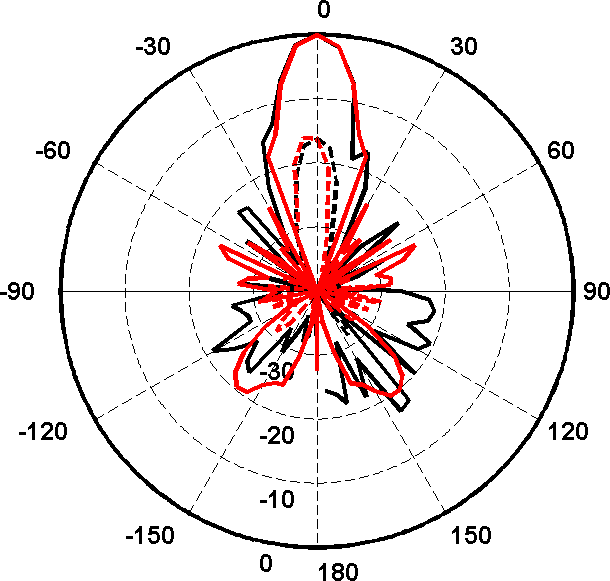}}}
	\mbox{\subfloat[]{\label{fig:DR_Nickel} \includegraphics[width=0.45\columnwidth]{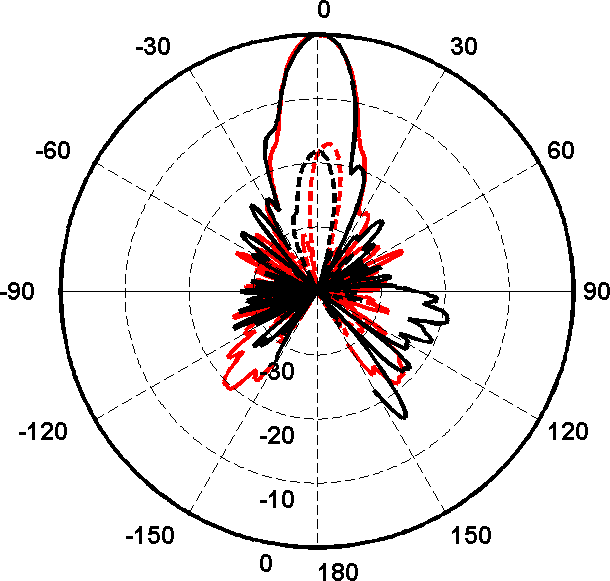}}}
	\caption{Simulated (a) and measured (b) radiation pattern of the reflector antenna. Both E-plane (black) and H-plane (red) co-polar (solid) and cross-polar (dashed) components are shown.}
	\label{fig:DR}
\end{figure}

The 3D radiation pattern was obtained in our anechoic chamber (Fig. \ref{fig:DR_Setup}) at $f_0=9.5$ GHz. Two planes, XZ plane ($\phi=0\degree$) and YZ plane ($\phi=90\degree$), and both co-polar and cross-polar components are shown in Fig. \ref{fig:DR_Nickel} for the nickel metalized reflector.  $\theta$ is swept between $\pm 150\degree$.

As expected, this design shows spill over: two symmetric secondary lobes at $\theta\approx\pm 150\degree$ in the XY plane (red, solid) and several ones for $\theta\geq90\degree$ (black, solid) are clearly visible. A further improvement of the main beam efficiency is also possible by reducing the feed angle in the YZ plane, which would reduce the secondary lobes for $\theta\geq90\degree$ but would increase the ones at $\theta\leq-90\degree$ and the losses due to the reflector blockage. Nevertheless, an SLL of 20 dB is achieved. The main beam is rotationally symmetric along the Z axis, which indicates a fairly homogeneous and symmetric reflector illumination.

\section{Conclusions}

An X band 3D printed reflector antenna has been described and characterized. It has been shown that this technology allows an easy, cost-affordable and highly flexible implementation of a relatively complex system, such as an off-axis parabolic reflector system. 

A novel dual polarization compact circular horn antenna with a DRW taper has been simulated and validated as a feed for the parabolic reflector. The whole antenna has a main beam rotationally symmetric and an SLL of 20 dB at its central frequency of $f_0=9.5$ GHz. An electromagnetic explanation  for this symmetrization has been given. Return losses are greater than 10 dB between 8 GHz to 11 GHz. It is lightweight (it weights less than 300 grams) and cost affordable.

\end{document}